\documentclass[11pt,aps,prd,nofootinbib,groupedaddress,preprintnumbers]{revtex4}
\usepackage{graphicx,epsfig}
\newcommand{\bea}{\begin{eqnarray}}
\newcommand{\beq}{\begin{equation}}
\newcommand{\eea}{\end{eqnarray}}
\newcommand{\eeq}{\end{equation}}

\newcommand{\lsim}{\raise0.3ex\hbox{$\;<$\kern-0.75em\raise-1.1ex\hbox{$\sim\;$}}}
\newcommand{\gsim}{\raise0.3ex\hbox{$\;>$\kern-0.75em\raise-1.1ex\hbox{$\sim\;$}}}

\newcommand{\unity}{{\hbox{1\kern-.8mm l}}}

\newcommand{ \overdot}{{\raise .2 ex \hbox to 0pt {\hss\bf\smash{.}\hss}}}
\newcommand{\bpm}{\begin{pmatrix}}
\newcommand{\epm}{\end{pmatrix}}

\def\ie{{\it i.e.}}

\begin{document}

\preprint{FTUV-08-0511}
\preprint{IFIC-08-69}

\title{Inflation might be caused by the right \tiny{(handed neutrino)}}
\author{G.~Barenboim}
\affiliation{Departament de F\'{\i}sica Te\`orica and IFIC, Universitat de 
Val\`encia-CSIC, E-46100, Burjassot, Spain.}

\begin{abstract}
We show that the scalar field that drives inflation can have a dynamical origin, being a
strongly coupled right handed neutrino condensate.
The resulting model is phenomenologically tightly constrained, and can be experimentally 
(dis)probed in the near future.  The  mass of the right handed neutrino 
obtained this way (a crucial ingredient
to obtain the right light neutrino spectrum within the see-saw mechanism in a complete
three generation framework) is related to that of the inflaton and both completely determine
the inflation features that can be tested by current and planned experiments. 
\end{abstract}

\maketitle

\section{Introduction}
Modern cosmology is built upon a theoretical framework whose foundations are
Big Bang theory and general relativity. Together they describe with 
precision many aspects of the universe from the first
nanosecond until our days. Despite its power, this
framework is unable to explain the observed flatness and homogeneity of space, let
alone the origin of matter and structure. As a consequence, this minimal
picture is usually supplemented by postulating
one or more episodes of accelerated expansion 
\cite{inflation} (inflation). 

The simplest realization of this idea is an approximately constant energy density, leading
to quasi-de-Sitter or exponential expansion. This can be parametrized
in terms of a fundamental scalar field, named the inflaton, whose nature is yet
unknown. Any other species that might have been present together with the inflaton
are quickly diluted away by the expansion, so that the inflaton is essentially 
playing ``solo'' during the inflationary epoch.

Although this idea is indeed attractive, fundamental scalars have not been observed yet.
Even more, since the rise and fall of the aether -- a primitive version
of a fundamental scalar -- alternative scenarios have been explored, in which
the scalar is an order parameter of some strong dynamics, rather than a
fundamental degree of freedom. In the famous 
BCS or Nambu-Jona-Lasinio mechanisms \cite{nambu} 
new interactions associated with a high energy scale $\Lambda$ are used to trigger
the formation of a low energy condensate, which mimics the role of a scalar.

This idea was followed by Bardeen, Hill and Lindner \cite{Bardeen:1989ds}, who proposed that 
a top quark condensate 
can replace the fundamental standard model Higgs boson to drive electroweak symmetry breaking.
A four fermion self-coupling of the top quark of strength $G$ signals the formation
of a top-antitop condensate, dynamically generating a mass for the top. Below the
cutoff scale $\Lambda$, one can integrate out the high frequency modes of the
fermions, obtaining an effective theory of a Higgs-like composite. In the large $N_c$
limit the theory predicts both the top mass and the scale of electroweak symmetry
breaking in terms of the fundamental parameters of the fermion theory at the
cutoff scale. The scale of electroweak symmetry breaking can be parametrically
small compared to $\Lambda$.

In this work, we will play the same game with the inflaton. We will attribute the 
dynamical origin of the inflaton field, to another ``solo'' player, 
the right handed neutrino \cite{hrn} . In analogy to the idea of top-quark condensation, we 
consider the standard model including right-handed neutrinos but without an inflaton. We add
four-fermion couplings which should be viewed as effective interactions that describe
the physics below a high energy cutoff, that we will take to be the Planck scale.
This new interaction should be strong
enough for a neutrino condensate that will trigger spontaneous symmetry breaking of
lepton number and produce a Majorana mass for the right-handed neutrino.
The same dynamics also produces ``natural inflation'' \cite{natural}. At the
same time that we obtain a slightly red spectral index, both the inflationary
de Sitter scale and the right-handed neutrino mass can naturally be of 
order $10^{17}$ GeV.
As compared to the usual approach to inflaton model building our dynamical
framework is both economical and predictive.

\section{Constructing the scalar field}
For simplicity we assume that one generation of right handed neutrinos 
has a four-fermion self-coupling,  
generated at a high energy scale $\Lambda$ from unknown dynamics. 
The underlying new dynamics might for example
be some non-abelian gauge interactions.
The four-fermion effective interaction for the right handed neutrinos below the scale 
 $\Lambda$ takes the form 
\beq
G \; \left( \bar{\nu}_R^c \; \nu_R \right) \left( \bar{\nu}_R \; \nu_R^c \right) 
\label{four-fermion}
\eeq
where $G$ is the dimensionful coupling constant, $\nu_R$ is the right handed neutrino and
$\nu^c$ indicates charge conjugation.
This is an effective interaction describing the physics below the physical
cutoff  $\Lambda$. There may be other higher dimension operators, but these will have
subdominant effects at energies substantially below the cutoff scale.

Analogously to \cite{Bardeen:1989ds} for a top condensate, 
the gap equation for a dynamically generated right-handed neutrino mass
has a solution when
\bea\label{eqn:gap}
G \Lambda^2 > \frac{8\pi^2}{N_f} \; , 
\eea
where $N_f$ is the number of right-handed neutrino flavors.
If the four-fermion interaction were the result of integrating out
a new non-abelian gauge interaction, $N_f$ could be the number of
``colors'' of this gauge theory. Strictly speaking, the analysis of
the condensate properties is performed in the limit of large $N_f$.

When  
the right-handed neutrinos condense
the condensation effects can be incorporated
by introducing an auxiliary scalar field $\Phi$ into the lagrangian
\beq
- m_o^2 \Phi^\dagger \Phi + g_o \left( \bar{\nu}_R^c \; \nu_R  \Phi + \mbox{h.c.} \right).
\eeq
Notice that the new effective scalar field does not have a kinetic term and reproduces the four-fermion
interaction when integrated out with
\beq
G=g_o^2/m_o^2.
\eeq

For the study of the low-energy regime we would like to keep the effective scalar field and integrate out
the short distance components of the right handed neutrino field. By doing so, we will see that at scales
below the cutoff, the effective scalar field develops fully gauge invariant induced kinetic terms and quartic 
self-interactions through fermion loops. The full induced effective lagrangian will take the form
\beq
g_o \; \left( \bar{\nu}_R^c \; \nu_R  \Phi + \mbox{h.c.} \right)\;\; + \;\;
Z_\Phi  \mid D_\mu \Phi \mid^2 \;\;- \;\; m_\Phi^2 \; \Phi^\dagger \Phi \;\;-\;\;
 \lambda_o \; \left(\Phi^\dagger \Phi \right)^2,
\eeq
where $D_\mu $ is the gauge covariant derivative, and all loops now to be defined with respect to a
low energy scale $\mu $ yielding for the induced parameters
\bea\label{eqn:zeqn}
Z_\Phi & = &\frac{ N_f \; g_o^2}{(4 \pi)^2}\;\; \ln\left(\frac{\Lambda^2}{\mu^2}\right) \\
&&\nonumber \\\label{eqn:meqn}
m_\Phi^2 & =& m_o^2\;\; - \;\; \frac{ 2 \; N_f \; g_o^2}{(4 \pi)^2}\;\; \left( \Lambda^2 -\mu^2 \right) \\
&&\nonumber \\\label{eqn:lameqn}
\lambda_o & = &\frac{ N_f \; g_o^4}{(4 \pi)^2}\;\; \ln\left(\frac{\Lambda^2}{\mu^2}\right). 
\eea
The mechanism for spontaneous symmetry breaking is now apparent 
in the effective scalar mass-squared, 
which is shifted by a negative finite value proportional to the cutoff $\Lambda $.

We would like to get a Lagrangian with a canonical kinetic term, which can be done by 
rescaling the scalar field $\Phi  \longrightarrow \Phi/\sqrt{Z_\Phi} $ and defining 
\bea
&& g = g_o/\sqrt{Z_\Phi} \\
&& m^2 =  m_\Phi^2 /Z_\Phi \\
&& \lambda  = \lambda_o/Z_\Phi^2,
\label{lambda}
\eea
to get
\bea
g \; \left( \bar{\nu}_R^c \; \nu_R  \Phi + \mbox{h.c.} \right)\;\; + \;\;
\mid D_\mu \Phi \mid^2 \;\; - \;\;  V(\Phi),
\eea
where $V(\Phi)$ is the scalar field potential and is given by 
\bea\label{eqn:effpota}
 V(\Phi) = \; m^2 \; \Phi^\dagger \Phi \;\;+\;\; \lambda \; \left(\Phi^\dagger \Phi \right)^2.
\eea
For $G$ satisfying (\ref{eqn:gap}), $m^2$ becomes negative at sufficiently low energies,
leading to spontaneous symmetry breaking with 
$\Phi$ developing a vev $v= \sqrt{-m^2/\lambda}$.

In the cosmological context we will want the analog of the effective potential (\ref{eqn:effpota})
at finite temperature; this is obtained by replacing (\ref{eqn:meqn}) by
\bea
m_\Phi^2 & =& m_o^2 + k^2T^2 - \;\; \frac{ 2 \; N_f \; g_o^2}{(4 \pi)^2}\;\; \left( \Lambda^2 -k^2T^2 \right)
\; ,
\eea
and similarly replacing $\mu$ by $kT$ in (\ref{eqn:zeqn}) and (\ref{eqn:lameqn}).
When $G$ satisfies (\ref{eqn:gap}), we can simplify the notation by defining
\bea
\delta &\equiv& \frac{N_fG\Lambda^2}{8\pi^2} - 1 \; , \\
\beta &\equiv& \frac{m_0}{\Lambda} \; .
\eea
At finite temperature the effective potential acquires a symmetry-breaking
minimum below the critical temperature given by
\bea
k^2T_c^2 =  \frac{\delta \beta^2 \Lambda^2}{1 + (1+\delta )\beta^2}
\; .
\eea
The vacuum expectation value of $\Phi$, which we denote by $v$, can be written as
\bea\label{eqn:phivev}
v^2 &=& \frac{\delta\beta^2\Lambda^2}{\lambda_0} \left( 1-\frac{T^2}{T_c^2} \right) \\
&=& \frac{g^2\delta \beta^2}{g_0^4} \Lambda^2 \left( 1-\frac{T^2}{T_c^2} \right)
\; .
\eea

\section{The phase field and its circumstances}
In the previous section we have generated a potential for the effective scalar field that represents
the physics of the neutrino condensate. For cosmology we will introduce by hand a cosmological
constant term so that $V( \Phi = \sqrt{-m^2/\lambda}) = 0$. With this addition, the potential can
now be written as 
\beq
V(\Phi) = \lambda \left( \Phi^\dagger \Phi - \tilde{m}^2 \right)^2
\eeq
with $ \;\lambda \; \tilde{m}^2 = m^2 $ and $\lambda \; \tilde{m}^4$ the cosmological constant term mentioned
before. Notice that this potential is invariant under a global $U(1)$ transformation $\Phi \longrightarrow 
e^{i \alpha} \Phi $, which is nothing less than lepton number.

Since $\Phi$ is a complex scalar field, it can be parametrized as 
$\Phi = \phi \;   e^{i \theta}$, and
the induced potential $V(\Phi)$ is a function of the radial field only, \ie\ 
\beq
 V(\Phi)\;=\;V(\phi) = \lambda \left( \phi^2 - \tilde{m}^2 \right)^2. 
\label{phi-4-potential}
\eeq
The radial field has a mass $m^2_\phi = \lambda  \tilde{m}^2 $. 

Potentials of the form 
(\ref{phi-4-potential}) are easy to analyze regarding inflation: in order to obtain sufficient
inflation (the famous 60 e-folds), the initial value of the field $\phi$ must be greater than the Planck mass.
In order to obtain acceptable density perturbations,
$\lambda$ must be about $10^{-15}$ \cite{lambda}. 
Such a small value for  $\lambda$ can be considered the 
kiss of death regarding the use of the radial field potential for
inflation. In our model, $\lambda$ is generated dynamically, not chosen by hand;
from (\ref{lambda}) it can be clearly seen that there is no $\Lambda - \mu$ combination
for which $\lambda$ can be so miserably small.

The phase field $\theta$ on the other hand, is a Nambu-Goldstone boson and is massless at tree level. 
If the $U(1)$ symmetry of the potential is preserved, the phase field will remain massless even with loop 
corrections. If the $U(1)$ is explicitly broken, the field $\theta$ will acquire a potential from loop corrections
leading to nonzero mass, becoming a pseudo Nambu-Goldstone boson. 
If this new mass is hierarchically smaller than that of the radial field, $\theta$ will be effectively massless near
the original symmetry breaking scale. When the temperature of the thermal bath drops below the mass
of the radial field, its excitations will be damped, and its expectation value rapidly settles into the
temperature-dependent minimum given by (\ref{eqn:phivev}). Thus for temperatures less than $O(T_c)$
the condensate dynamics is described by $\theta$ alone. Its potential  
$V_\theta (\theta)$, which was negligible for high temperatures, becomes important
below $T_c$. This phase field then rolls down its
potential to its minimum, acting as the driver for an extended period of inflation.

To compute the outcome of inflation driven by a neutrino condensate, we first need
to specify the explicit breaking of the $U(1)$ symmetry from dynamics above the
cutoff scale.
On general grounds it is expected that global symmetries such as our lepton number might be broken explicitly
by Planck scale physics \cite{Holman:1992us}. 
Two general arguments support this assertion. The first comes along the black-hole ``no hair''
theorems of classical relativity: as black-holes cannot support any global ``hair'', a virtual exchange of black-holes 
would give rise to global symmetry violating operators in the low energy theory, where low energy in this case
means energies smaller than the mass of the black-hole.
The second argument arises courtesy of the existence of wormhole configurations \cite{Giddings:1988cx}. 
While gauge charges cannot be sent down
a wormhole there is nothing to prevent the loss of global charges this way \cite{Gilbert:1989nq}. 
The effective theory at scales below the
wormhole scale will have non-zero global charge carrying operators. Integrating out the effective black holes will
bring Planck scale suppressed non-renormalizable operators that break the global $U(1)$.

Thus if we take the cutoff scale to be the Planck scale, $\Lambda \simeq 10^{19}$ GeV,
it is natural to expect an explicit breaking of the global $U(1)$. The lowest dimension
symmetry-breaking operator constructed from the right-handed neutrinos is given by
\beq
G^\prime  \;\left[ \left( \bar{\nu}_R^c \; \nu_R \right)^2 +  \left( \bar{\nu}_R \; \nu_R^c \right)^2 \right].  
\label{lfv-four-fermion}
\eeq
As we have seen before, we can reproduce the physics of the four-fermion coupling by resorting to the scalar field 
$\Phi$ as before, adding the interaction 
\beq
g^\prime \; \left( \bar{\nu}_R^c \; \nu_R  \; \; \Phi^\dagger  \; \; +  \;
 \; \bar{\nu}_R \; \nu_R^c   \; \;\Phi  \right). 
\eeq
Under a global phase redefinition, 
\bea
&&\nu_R \longrightarrow e^{i \alpha} \nu_R \nonumber \\
&&\nu_R^c \longrightarrow e^{-i \alpha} \nu_R^c \nonumber \\
&&\Phi \longrightarrow e^{2 i \alpha} \Phi \nonumber
\eea
the new term is not invariant, explicitly breaking the $U(1)$ 
symmetry down to a residual discrete symmetry generated by
$\theta \to \theta +2\pi $. This results in a nontrivial potential for the
effective pseudo Nambu-Goldstone
boson from the condensate. This field will play the role of the inflaton in our model.

Neglecting the shift in the vacuum expectation value of the radial field induced at 1-loop, the tree level 
right handed neutrino mass will now be given by 
\bea
m_R^2 (\theta) = (g^2 + g^{\prime 2}  + 2gg^\prime \cos(\theta )  ) v^2
\;,
\eea
with $v^2$ given by (\ref{eqn:phivev}), in the appoximation that $g^\prime \ll g$.
The explicitly broken $U(1)$ symmetry is reflected in the mass dependence of the right handed neutrino mass
on the phase field $\theta $. Quantum effects will produce a potential for this field, thus providing 
our inflationary 
potential. At the 1-loop level this is given by 
\bea
V_\theta (\theta) \;&= &\;-\frac{1}{(16 \pi)^2} \;\left( m^2_R (\theta) \right)^2 \;
\ln\left(\frac{m^2_R (\theta)}{v^2}\right) \nonumber \\
\nonumber \\
&=&\;- \;\frac{g^2 g^{\prime 2} v^4}{16 \pi^2} \; \left[ \frac{g^2 + g^{\prime 2}}{2 g g^\prime} + 
\cos(\theta ) \right]^2 
\;\ln \left[ 2 g g^\prime \left( \frac{g^2 + g^{\prime 2}}{2 g g^\prime} + \cos(\theta ) \right) \right].
\eea
This potential has extrema  at $\theta  =\; 0, \pi,\; \cos^{-1}\left( -\frac{g^2 + g^{\prime 2}}{2 g g^\prime} + 
\frac{1}{2 g g^\prime \sqrt{e}} \right) $ being $\theta = \pi$ the only  minimum.

The mass of the pseudo Nambu-Goldstone boson at the minimum of the potential, \ie\  the mass of the inflaton
is given by
\bea
 m_\theta^2 \equiv \left. \frac{\partial^2  V_\theta}{\partial \theta^2} \;  \right|_{\theta = \pi}  = 
- \;\frac{g  g^{\prime } v^2}{32 \pi^2 } \; \left( g - g^\prime \right)^2 \; \; 
\left[ 1 + 2 \ln\left(\left(g - g^\prime \right)^2 \right) \right]
\eea
while that of the right handed neutrino reads
\bea
\left. m_R^2 \; \right|_{\theta = \pi}  \; = \; \left(g - g^\prime \right)^2  \; v^2
\eea
Notice that the potential as well 
as both masses are invariant under the exchange $g \longleftrightarrow g^\prime$ and
that both masses vanish when $g = g^\prime$; this corresponds to the degenerate case
that only the real part of $\Phi$ couples to the neutrinos, meaning that the
condensate is uncharged.

As in the radial field case, the potential for the phase field does not vanish at its minimum,
a fact that we are going to change by defining a new phase field potential given by
\bea
V(\theta) &=& V_\theta (\theta) - V_\theta( \theta = \pi) \nonumber \\\nonumber \\
&=& \;- \;\frac{g^2 g^{\prime 2} v^4}{16 \pi^2} 
\; \left\{  
\left[ \frac{g^2 + g^{\prime 2}}{2 g g^\prime} + 
\cos(\theta ) \right]^2 
\;\ln \left[ 2 g g^\prime \left( \frac{g^2 + g^{\prime 2}}{2 g g^\prime} + \cos(\theta ) \right) \right] 
\right. \nonumber \\
&& \left. -  \left[ \frac{\left(g^2 - g^{\prime 2}\right)^2}{2 g g^\prime} \right]^2 
\;\ln \left[ \left( g - g^{\prime}\right)^2  \right] \right\}.
\eea
On general grounds, one would expect $g^\prime \ll g $, \ie\ small explicit breaking. In this case, the effective potential
takes the delightfully simple form
\bea
V (\theta)  \;\simeq \; - \; \frac{g^3 g^{\prime } v^4}{32 \pi^2} \; \left[ 1 + 2 \ln\left(g^2\right) \right] 
\; \left( 1 + \cos\left(\theta \right) \right)
\eea
which is of the form of the well-known natural inflation potential \cite{natural} 
$ M^4 \left( 1 +  \cos\left(\theta \right) \right)$
with 
\bea
M^4 =- \; \frac{g^3 g^{\prime } v^4}{32 \pi^2} \; \left[ 1 + 2 \ln\left(g^2\right) \right]. 
\eea
Notice that although {\it prima facie} this constant term looks negative, it is indeed positive for $g <1$.

\section{Inflation phenomenology}
Detailed analysis of the virtues of natural inflation models already exist in the literature \cite{Freese:2008if}.
Here we will briefly review the basic features of our model and 
derive the constraints that available data impose on the
model parameters.

For an inflationary theory to correspond to the observed universe, it must satisfy at least two conditions :
(i) it has to explain the observed thermal equilibrium of the cosmic microwave background radiation (CMB)
which is guaranteed by 
providing sufficient inflation, \ie\ the inflationary potential  must drive an increase on the scale factor
of a minimum of $e^{60} $; (ii) the quantum fluctuations of the inflaton should give rise to primordial
density fluctuations of size $\delta \rho/\rho$ and spectral index $n_s$ in agreement with observations \cite{Komatsu:2008hk}.

During inflation,  the inflaton rolls down towards the minimum of its potential, evolving according to
\beq
\ddot{\theta} + 3 H \dot{\theta} + \partial V/\partial\theta =0
\eeq
where the Hubble rate $H$ is given by 
\beq
H^2\; = \;\frac{8 \pi}{3 m_{\mbox{Pl}}^2} \; \left[ \frac{\dot{\theta}^2}{2} + V(\theta) \right].
\eeq
In general, we are interested in potentials which contain one region flat enough that the evolution of the
field is friction dominated, (what goes under the name of slow-roll approximation) so that the equation of motion
is essentially given by 
\beq
3 H \dot{\theta} + \partial V/\partial\theta =0.
\eeq
Within the slow-roll approximation, the number of e-folds of inflation when the field evolves from $\theta$
to $\theta_f$ is 
\bea
N(\theta) =  \;\frac{8 \pi}{3 m_{\mbox{Pl}}^2} \; \int_{\theta_f}^\theta \frac{V(\theta)}{V^\prime(\theta)} d\theta
\eea
where $V^\prime(\theta) = \partial V/\partial\theta $ and $\theta_f$ is the value of the field at which
inflation stops (reheating commences) and is obtained from
\beq
\epsilon(\theta_f) \equiv \frac{ m_{\mbox{Pl}}^2}{16 \pi} \left[ \frac{V^\prime(\theta_f)}{V(\theta_f)} \right]^2 = 1
\eeq 
An upper limit on the initial value of the field  is obtained by imposing $N(\theta_i) = 60 $. Around this value,
quantum fluctuations on scales observed today were produced and its size is given by \cite{Lidsey:1995np}
\bea
\delta\rho/\rho \; \simeq \; \frac{ \left(\;V(\theta_i)\;\right)^{3/2}}{m_{\mbox{Pl}}^3 V^\prime(\theta_i)}
\eea
The spectral index of these density perturbations and its dependence on the scale read
\bea
n_s - 1  \; = \; - 6 \epsilon(\theta_i) + 2 \eta(\theta_i)  \\
\nonumber \\
\frac{d n_s}{d \ln k} = -16 \epsilon(\theta_i) \eta(\theta_i) + 24 \epsilon(\theta_i)^2 + 2 \xi^2(\theta_i)
\eea
where $\eta$ and $\xi^2 $ are the second  and third slow-roll parameters
\bea
\eta \; =\;  \frac{m_{\mbox{Pl}}^2}{8 \pi}\;\; \frac{V^{\prime \prime}}{V}   \;\; \;\; \mbox{and} \;\; \;\; 
\xi^2 \; =\;  \frac{m_{\mbox{Pl}}^4}{(8 \pi)^2 }\;\;\frac{ V^{\prime \prime} \; V^\prime}{V^2}. 
\eea 
Expressed in terms of the parameters of the model, these quantities are given by
\bea
&&\sin(\theta_i )= \left[ \beta \left(2 -\beta \right)^2 \right]^{1/2} \\
&&\nonumber \\
&&\delta\rho/\rho \; \simeq \; \frac{M^2 v}{m_{\mbox{Pl}}^3} \; \frac{\left(2 -\beta \right)}{\beta^{1/2}} \\
&&\nonumber \\
&&\epsilon =  \frac{m_{\mbox{Pl}}^2}{16 \pi v^2} \; \frac{\beta}{\left(2 -\beta \right)} \\
&&\nonumber \\
&&\eta =  \frac{m_{\mbox{Pl}}^2}{8 \pi v^2} \;\frac{\left(1 -\beta \right) }{\left(2 -\beta \right)}\\
\nonumber \\
&& n_s - 1 =  \frac{m_{\mbox{Pl}}^2}{16 \pi v^2} \;\frac{2 \left(2+\beta \right) }{\left(2 -\beta \right)} \\
&&\nonumber \\
&&\xi^2 = -\frac{m_{\mbox{Pl}}^4}{(8 \pi)^2  v^4} \;\frac{\beta^{1/2} \left(1-\beta \right) }{\left(2 -\beta \right)^{3/2}}
\eea
where
\bea
\beta = \frac{2 \; e^{-2 N y}}{y (1+y)}  \;\; \;\; \mbox{with} \;\; \;\; y = \frac{m_{\mbox{Pl}}^2}{16 \pi v^2} 
\eea
and $N$ number of efolds before the end of inflation at which observable perturbations were
generated.  

\begin{figure}

\centerline{\epsfxsize 3.75 truein \epsfbox {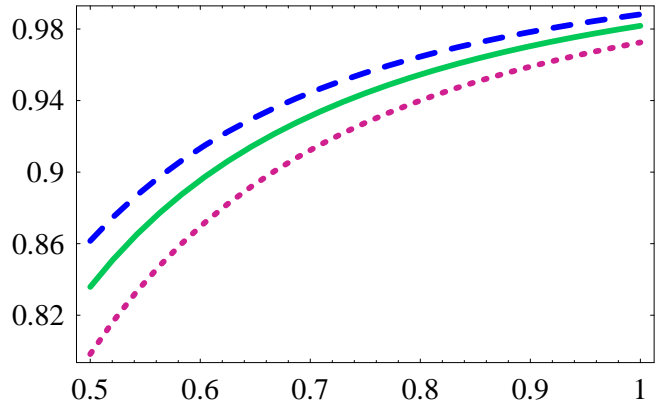} }
 
\caption{Spectral index of the complete potential 
as a function of the symmetry breaking scale $v$ for different values of the couplings }
\begin{center}
\begin{small}   
\vspace{-1cm}
\begin{tabular}{| l | c | r | }
    \hline
      & $g$ $\;$ $\;$  & $g^\prime $ $\;\;$ $\;\;$ \\ \hline
     green (solid)  & $\;\; $ 0.01  $\;\;$ &  $\;\;$ 0.001 $\;\;$ \\ \hline
     blue (dashed) & 0.01  &$\;\;$  0.0001$\;\;$  \\ \hline
     purple (dotted) &0.1 & $\;\;$ 0.03 $\;\;$  \\
    \hline
  \end{tabular}
\end{small}
\end{center}
\label{fig:ns}
\end{figure}

These simplified expressions, although  very easy to work with, hide the dependence
of the cosmological observables on the model parameters. This becomes apparent in the fact that all the 
quantities above, but $\delta\rho/\rho $, depend exclusively on $v $. In order to link the original model
parameters, $v$, $g$ and $g^\prime$ to observations, the full potential must be used. Nevertheless, the slow-roll 
approximation is still reasonable.

As can be seen from Figure(\ref{fig:ns}), the spectral index of the density fluctuations defines the range
of values the parameter $v$ can take.  Clearly, although both coupling do contribute to the value
of the spectral index, cosmological observations force  $v$ to live in the range $  \; 
0.7 \, m_{\mbox{Pl}} \; < \; v \; < \; 
 0.9 \, m_{\mbox{Pl}}  \;$ for any reasonable choice of  $g$ and $g^\prime$.  
In this figure the spectral index is evaluated at a value $\theta_i$ such that
sufficient inflation occurs when the field rolls down from $\theta_i$ to the end of inflation.

\begin{figure}

\centerline{\epsfxsize 3.75 truein \epsfbox {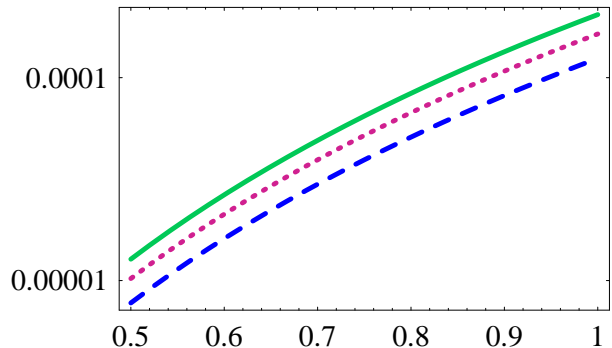} }
 
\caption{Magnitude of the density fluctuations
as a function of the symmetry breaking scale $v$ for different values of the couplings }
\begin{center}
\begin{small}   
\vspace{-1cm}
\begin{tabular}{| l | c | r | }
    \hline
      & $g$ $\;$ $\;$  & $g^\prime $ $\;\;$ $\;\;$ \\ \hline
     green (solid)  & $\;\; $ 0.01  $\;\;$ &  $\;\;$ 0.001 $\;\;$ \\ \hline
     blue (dashed) & 0.01  &$\;\;$  0.0003$\;\;$  \\ \hline
     purple (dotted) &0.1 & $\;\;$ 0.00001 $$  \\
    \hline
  \end{tabular}
\end{small}
\end{center}
\label{fig:delta}
\end{figure}

Once the symmetry breaking scale is defined to take values within this range, we can resort to 
the magnitude of density fluctuations
to see whether some information on the couplings can be obtained. The answer is depicted on Figure(\ref{fig:delta}): 
the size of the primordial perturbations, evaluated at a value $\theta_i$ such that
sufficient inflation occurs, does strongly depend on the couplings. However, it depends through the
combination $( \; g^3 g^\prime \; )^{1/2}$, so only this combination can 
be bounded using the magnitude of density perturbations, and it 
must be $( \; g^3 g^\prime \; )^{1/2} \;  \sim  \; 10^{-5}$.
It is important to notice that the scale $v$ and the coupling $g$ determine not only the
the mass
of the inflaton (provided $g^\prime \ll g $) but also  the 
mass of the right-handed neutrino and the scale of spontaneous symmetry breaking; 
thus the importance of connecting both values 
with cosmological observations.

As an example, here is a set of input parameters from the cutoff scale that
will produce a satisfactory model of inflation:
\bea
\Lambda = m_0 = 10^{19}{\rm\ GeV}\; ;\quad G\Lambda^2 = 0.1 \; ;\quad \frac{8\pi^2}{N_f} = 0.05
\; .
\eea
From which we derive $\delta = 1$ and $g \sim 0.1$. We then require $g^\prime \sim 10^{-7}$
to get the right magnitude of density perturbations. In this example the symmetry
breaking scale $v$ is close to the Planck scale, while $m_R$ and $M$ are of order $10^{18}$
GeV. Escaping from the large $N_f$ limit assumed here would require a better understanding
of the dynamics responsible for the condensate.

To bound the value of the couplings $g$ and $g^\prime$ separately, a different observable with a 
different dependence
on the couplings needs to be found. Two possibilities
are at hand, although neither will be measured any time soon.  Nevertheless, future measurements
can rule out a dynamical origin of the inflaton field, as the one proposed here.

\begin{figure}

\centerline{\epsfxsize 4.75 truein \epsfbox {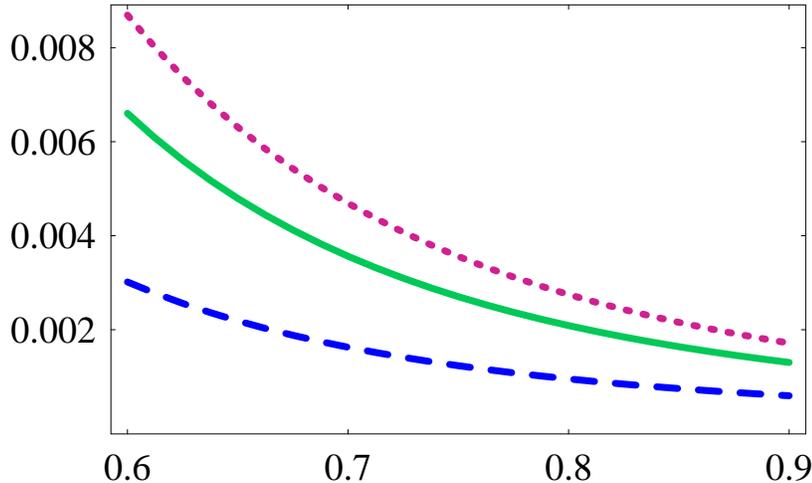} }

\caption{Tensor to scalar ratio 
as a function of the symmetry breaking scale $v$ for different values of the couplings }
\begin{center}
\begin{small}   
\vspace{-1cm}
\begin{tabular}{| l | c | r | }
    \hline
      & $g$ $\;$ $\;$  & $g^\prime $ $\;\;$ $\;\;$ \\ \hline
     green (solid)  & $\;\; $ 0.001  $\;\;$ &  $\;\;$ 0.003 $\;\;$ \\ \hline
     blue (dashed) & 0.1  &$\;\;$  0.000003$\;\;$  \\ \hline
     purple (dotted) &0.01 & $\;\;$ 0.005 $$  \\
    \hline
  \end{tabular}
\end{small}
\end{center}
\label{fig:16epsilon}
\end{figure}

In addition to scalar (density) perturbations, our field will also give rise to tensor (gravitational wave) 
perturbations. Generally, the tensor amplitude is given in terms of the tensor/scalar ratio 
\beq
r \equiv \frac{P_T}{P_R} = 16 \epsilon 
\eeq
which is shown in Fig(\ref{fig:16epsilon}). The tensor to scalar ratio $r$ goes like  $\; g^2 g^{\prime 2} \; $,
and would offer the possibility of bounding each coupling individually, if the tensor amplitude were not well
below the detection sensitivity of current and (near) future experiments, \ie\  gravity waves are exponentially 
suppressed relative to the adiabatic scalar fluctuations over the observable large scale waveband. 
Gravity waves are the holy grail of next generation of experiments \cite{next} 
and if found, will rule out this model.

Notice that in order to describe scalar an tensor fluctuations, only four parameters are needed (if we ignore
the running): the amplitude and the spectra of both modes. The spectral indexes $n_s$ and $n_T \equiv - 2 \epsilon$ characterize
the latter, while the size of the scalar perturbations is basically characterized by the height of the
potential (given by $M^4$ in the approximated expressions). The tensor amplitude is given by $r$.
However, the tensor index is not and independent parameter since its related to the tensor/scalar ratio
by the inflationary consistency relation  $r = - 8 n_T $ and therefore it is not useful for disentangling
the values of each coupling.

\begin{figure}

\centerline{\epsfxsize 4.75 truein \epsfbox {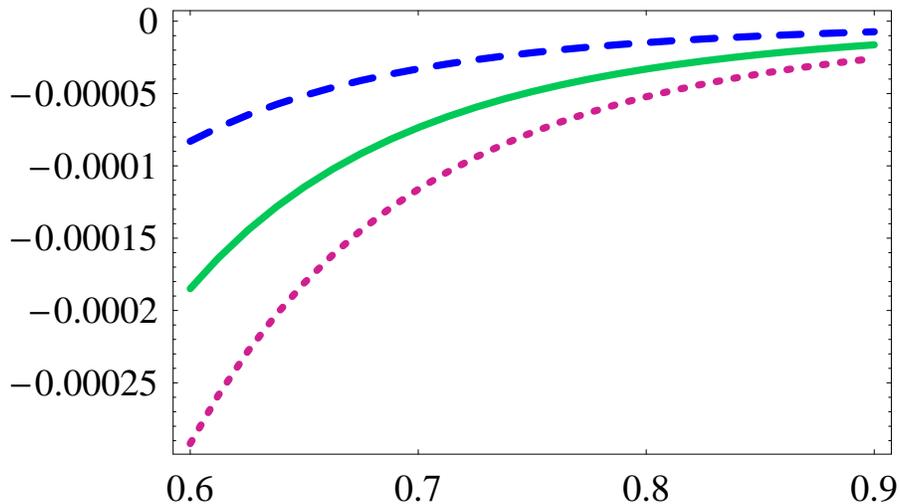} }
 
\caption{Running of the spectral index  
as a function of the symmetry breaking scale $v$ for different values of the couplings }
\begin{center}
\begin{small}   
\vspace{-1cm}
\begin{tabular}{| l | c | r | }
    \hline
      & $g$ $\;$ $\;$  & $g^\prime $ $\;\;$ $\;\;$ \\ \hline
     green (solid)  & $\;\; $ 0.01  $\;\;$ &  $\;\;$ 0.003 $\;\;$ \\ \hline
     blue (dashed) & 0.1  &$\;\;$  0.000003$\;\;$  \\ \hline
     purple (dotted) &0.01 & $\;\;$ 0.005 $$  \\
    \hline
  \end{tabular}
\end{small}
\end{center}
\label{fig:dndk}
\end{figure}

In general, $n_s$ is not a constant, and its dependence on the scale can be characterized by its running. 
Unfortunately the slow-roll approximation is numerically inaccurate for this parameter and may lead
to discrepancies of a factor 2-3. However, we are interested in the order of magnitude of the result
and therefore using the slow-roll approximation will leave our conclusions unaffected. 
As shown in Fig(\ref{fig:dndk}) our model predicts a
very small and negative spectral index running, scaling as $g^\prime / g$. 
It is so negligible small  that it is essentially indistinguishable
from zero running. Small scale CMB experiments \cite{smallscale}
 will provide more stringent tests on the running. If these
experiments exclude a trivial (consistent with zero)  running, \ie\ if they detect a strong running, our model would
be ruled out.

\section{Conclusions}
We have shown that the scalar field that drives inflation can have a dynamical origin, being a
strongly coupled right handed neutrino condensate. The fact that $\Phi$ behaves like a sensible
propagating field is a signal that we have chosen the correct low-energy degrees of freedom by introducing it.
As the theory containing  $\Phi$  is equivalent to a theory entirely written in terms of neutrino degrees
of freedom, the field $\Phi$ can be interpreted as a right handed neutrino bound state.

The resulting model is phenomenologically tightly constrained, and can be experimentally 
(dis)probed in the near future. Probably the least attractive feature of the model is the range
of values the symmetry breaking scale $v$ is bounded to take, quite close to the Plank scale.
This won't be the case in a scenario with more than one generation of right handed neutrinos.
We wish to emphasize however, that the mass of the right handed neutrino (a crucial ingredient
to obtain the right light neutrino spectrum within the see-saw mechanism in a complete
three generation framework) is related to that of the inflaton and both completely determine
the inflation features that can be tested by current and planned experiments. 
Thus, despite its problems we feel that the proposed dynamical origin of the inflaton field is
sufficiently interesting to merit attention.

\section*{Acknowledgments}
My warmest thanks to Bill Bardeen for his expert advice, encouragement and 
support.
I am also grateful to Joe Lykken for his invaluable help and to
 Oscar Vives  for useful discussions and acknowledge support from the Spanish MEC and FEDER
under  Contract FPA2005-01678 and the Generalitat Valenciana GVPROMETEO2008-004.

\end{document}